\documentstyle[twoside,fleqn,espcrc2]{article}	 

\def\BUCKOW{	\par\vspace{7pt} 
		Talk at the \it 30th International Symposium Ahrenshoop on the
		Theory of Elementary Particles \rm
		in Buckow, August 27-31,
		1996\\[-20pt]\phantom.
}

\catcode`\"=\active \let"=\"

\def\ifundefined#1{\expandafter\ifx\csname#1\endcsname\relax}

\def\HS#1 {\hspace*{#1pt}} \def\VS#1 {\vspace*{#1pt}} \long\def\del#1\enddel{} 
\def\BC{\begin{center}}	   \def\VR#1#2{\vrule height #1mm depth #2mm width 0pt}
\def\EC{\end{center}}	   \def\TVR#1#2{@{~~\VR{#1}{#2}}}	
\def\BP{\begin{picture}}   \def\EP{\end{picture}}

\let\0=\over	  \let\1=\vec	   \def\2{{1\over2}}	\let\3=\ss
\let\4=\underline \let\5=\overline \let\6=\partial	\def\7#1{{#1}\llap{/}}
\def\8#1{{\textstyle{#1}}}         \def\9#1{{\ifmmode{\pmb{#1}}\else\bf#1\fi}}

\def\({\left(}       \def\){\right)}	\def\Id{{\91}}	  

\def\EEL#1 {\label{#1}\EE}		
\def\BE {\begin{equation}} 	\def\EE {\end{equation}}	
\def\BEA{\begin{eqnarray}}	\def\EEA{\end{eqnarray}} 

\ifundefined{Bbb}
\font\bbfS=msbm7  \font\bbfSS=msbm5	\newfont\bbfT{msbm10 scaled 1000}
\newfam\BBFF	\textfont\BBFF=\bbfT	\scriptfont\BBFF=\bbfS	
\def\Bbb#1{{\fam\BBFF\relax#1}}		\scriptscriptfont\BBFF=\bbfSS	\fi
\def\pmb#1{\setbox0=\hbox{${#1}$}   \kern-.025em\copy0\kern-\wd0
      \kern.05em\copy0\kern-\wd0     \kern-.025em\raise.0433em\box0 }
   
  \def\IZ{{\Bbb Z}}  

\def\mao#1{\mathop{\rm #1}\nolimits}	
		
\def\mod{\mao{mod}}		

\let\and=\wedge		 		\let\ex=\times

\let\bra=\langle	\let\ket=\rangle	\def\<#1\>{\bra #1 \ket}

\def\rel#1 #2{\buildrel #1 \over {#2}}	

\let\a=\alpha			\let\d=\delta 	
 	 	 	
		\let\m=\mu	
	 	\let\p=\pi 		\let\s=\sigma 
 	 		 	
\let\Ph=\phi 	 	 	 	 
	 	 	 	\let\D=\Delta

\def\ca{{\cal A}}   
  \def\cg{{\cal G}}

\def\plb#1 #2 {Phys. Lett. {\bf B#1} #2 }
\def\phr#1 #2 {Phys. Rep. {\bf  #1} #2 }	
\def\npb#1 #2 {Nucl. Phys. {\bf B#1} #2 }
\def\aph#1 #2 {Ann. Phys. {\bf #1} #2 }		
\def\jmp#1 #2 {J. Math. Phys. {\bf #1} #2 }
\def\jgp#1 #2 {J. Geom. Phys. {\bf #1} #2 }
\def\prd#1 #2 {Phys. Rev. {\bf D#1} #2 }
\def\prl#1 #2 {Phys. Rev. Lett. {\bf #1} #2 }
\def\rmp#1 #2 {Rev. Mod. Phys.  {\bf #1} #2 }
\def\zpc#1 {Z. Phys. {\bf #1C} }
\def\cmp#1 #2 {Commun. Math. Phys. {\bf #1} #2 }
\def\cqg#1 #2 {Class.Quant.Grav. {\bf #1} #2 }
\def\mpl#1 {Mod. Phys. Lett. {\bf A#1} }
\def\cpc#1 {Computer Phys. Commun. {\bf #1} }	
\def\ijmp#1 {Int. J. Mod. Phys. {\bf A#1} }
\def\ijmpC#1 {Int. J. Mod. Phys. {\bf C#1} }

\newcounter{TRefNX} \let\OLDcite=\cite	\makeatletter
\def\makeTRefs#1{\@for 	\NewTRef:=#1\do{\global\makeTRef{\NewTRef}}}
\def\makeTRef#1{\ifundefined{TRef#1}\stepcounter{TRefNX}%
\expandafter\xdef\csname TRef#1\endcsname{\theTRefNX}\fi}\makeatother
\def\NEWcite#1{\makeTRefs{#1}\OLDcite{#1}}  

\ifundefined{draftmode} {} \else	\let\cite=\NEWcite
\newcount\HOUR 	\HOUR=\the\time \divide\HOUR by 60	\multiply \HOUR by 60 
\newcount\MIN	\MIN=\the\time 	\advance\MIN by -\HOUR 	\divide\HOUR by 60
\ifnum	\MIN>9	\def\printTIME{{\it\the\HOUR\,:\,\the\MIN}}
	\else	\def\printTIME{{\it\the\HOUR\,:\,0\the\MIN}} \fi 

\def\LLab#1{\BP(0,0)\unitlength=1mm\put(-9,.5){\makebox(0,0)[cr]{\tiny #1
        \rlap{$_{_{\makeatletter\csname TRef#1\endcsname\makeatother}}$}}}\EP}
\fi

\def\bye{\end{document}}		\long\def\new#1\endnew{{\bf #1}}

\title{ {\LARGE\bf (0,2) string compactifications }
							\hfill {\normalsize 
	TUW--96/22\BP(0,0)(0,0)\put(0,17){\llap{hep-th/9611130}}\EP	}}

\author{M.Kreuzer\thanks{e-mail: kreuzer@tph16.tuwien.ac.at} and
        M. Nikbakht--Tehrani\thanks{e-mail: nikbakht@tph16.tuwien.ac.at}
       \address{Institut f\"{u}r Theoretische Physik, 
                Technische Universit\"{a}t Wien\\
                Wiedner Hauptstra\3e 8--10, A-1040 Wien, AUSTRIA}}

\begin{document}

\begin{abstract}
Using the simple current method we study a class of $(0,2)$ SCFTs which we
conjecture to be equivalent to $(0,2)$ $\sigma$ models constructed 
in the framework of gauged linear sigma models.
					 			\BUCKOW
\end{abstract}

\maketitle

\section{Introduction}

It is well-known \cite{ba88} that the classical string solutions with
$N=1$ spacetime supersymmetry require the internal CFTs to have 
$(0,2)$ worldsheet supersymmetry. $(0,2)$ models lead to
more realistic  gauge groups like $SO(10)$ or $SU(5)$ \cite{wi86},
and are therefore phenomenologically more interesting 
than the $(2,2)$ case. To understand the space of ground states of classical
string theory there is little reason to restrict oneself to the 
study of $(2,2)$ models. But the assertion in \cite{di86}
that the generic $(0,2)$ $\sigma$ model is destabilized by worldsheet 
instantons has slowed down 
further development in this direction.
The revival of $(0,2)$ $\sigma$ models has been initiated mainly by the work 
of Distler and Greene \cite{di871,di881}, who gave a criterion for evading 
destabilization of the vacuum. 
Its verification is, however, not trivial. 
More recently Distler and Kachru \cite{di94} constructed a 
large class of $(0,2)$ $\sigma$ models 
that uses Witten's gauged linear sigma model approach \cite{wi93}. 
The resulting class of models is now believed to define appropriate 
SCFTs at the infrared fixed point \cite{si95}.

On the side of exact conformal models the situation seems to be more 
transparent. Using simple current techniques Schellekens and
Yankielowicz \cite{sc90} produced a telephone book of $(1,2)$ models from
tensor products of minimal models \cite{sc90T}. 
The main problem in this context is the 
arbitrariness in the selection of a managable 
subset of the models that are accessible with the known constructions
\cite{fo89,sc90}. 
This effort culminated in Schellekens' theorem
on the conditions for the possibility of avoiding fractional electric charges
\cite{sc90e,ly96}. 

An interesting question is, of course, to what extent one can identify 
the models constructed by these two approaches. Searching for such a 
class of models we will try to generalize the proposal of  
\cite{bl95,bl96}, which was originally made on the basis of a 
stochastic computer search 
for matching particle spectra. We will extend this to a series 
of identifications based on an analysis of the anomaly cancellation
conditions and on the assumption that the ${\Bbb Z}_4$ breaking mechanism 
(see below) does not care about the rest of the conformal field theory and
only acts on a Fermat factor of a 
non-degenerate Landau--Ginzburg superpotential.
This provides us with 3219  $SO(10)$ models, 
based on the list of 7555 weights for transverse hypersurfaces
in weighted projective spaces \cite{nms,kl94}, and many more if we include 
orbifolding or hybrid constructions.

After a quick review of $(0,2)$ compactification in the geometric
context in section 2 
we start with an ansatz to solve the anomaly cancellation equation and 
then briefly discuss some issues concerning the toric resolution of 
singularities \cite{bo93,ba95,di96}. 
In section 3 we discuss some
aspects of exact constructions in the framework of simple currents and
apply these to the class of models suggested by the $\sigma$ model connection.
We conclude with some comments about open problems and directions for future
work.

\section{$(0,2)$ $\sigma$ models and anomaly cancellation conditions}      
 
The geometric data that define a $(0,2)$ compactification \cite{hu85,wi86}
consist of
a K\"{a}hler manifold $M$ with K\"{a}hler form $J$ and some holomorphic
vector bundle
$V$ of rank r. The right moving fermions on the worldsheet couple to
the pull back of the tangent bundle $TM$ along the sigma model map and the 
left moving ones couple to the pull back of $V$. 
The very existence of these 
spinors requires that the topological obstruction to their existence 
vanishes, i.e. $c_{1}(V)=c_{1}(M)= 0$ (mod 2)\footnote{For a complex vector 
bundle the mod 2 reduction of $c_{1}$ is the second Stiefel--Whitney class.}.
Anomaly cancellation imposes a further condition on these data, 
namely $ch_{2}(M)=ch_{2}(V)$, where $ch_{2}=\frac{1}{2}c_{1}^{2}-c_{2}$
is the second Chern character. 
Vanishing of the lowest order beta functions requires that the K\"{a}hler
metric is Ricci-flat, i.e. $M$ is a Calabi-Yau manifold, and the 
connection on $V$ has to satisfy the conditions
$F_{ij} = F_{\bar{\imath}\bar{\jmath}}=0$ (i.e. it 
is holomorphic) and $g^{i\bar{\jmath}}F_{i\bar{\jmath}}=0$.
Due to the results of Donaldson and Uhlenbeck and Yau \cite{wi86,uh86} we 
have the following information about the set of solutions of $g^{i\bar{\jmath}}
F_{i\bar{\jmath}}=0$: If 
	the integrability condition $\int_{M}J^{n-1}\wedge c_{1}(V)=0$ 
is satisfied then the existence and uniqueness of the solution
is equivalent to the stability of $V$ with respect to $J$ on $M$. 
The stability of $V$ means that
for every `coherent subsheaf' $\cal F$ of $V$ with $0<\mao{rank}({\cal F})<
\mao{rank} (V)$ one has $\mu({\cal F})<\mu( V)$, where $\mu $, called
normalized first Chern number or slope, is defined by
$\mu({\cal F}):=\frac{1}{\mao{rank}({\cal F})}\int_{M}J^{n-1}
\wedge c_{1}({\cal F})$.

\subsection{The gauged linear sigma models}

It is believed that a `generic' $\sigma$ model of the type discussed above 
flows to a $(0,2)$ string vacuum in the infrared limit \cite{wi86,di881}.  
As pointed out in \cite{di881}, such models evade destabilization by 
world sheet instantons \cite{di86} if the splitting type of $V$ 
on all instantons in the Calabi-Yau space $M$ is nontrivial. 
However, it is technically difficult to check this condition.
A real breakthrough has been achieved  
by Witten's gauged linear sigma model approach \cite{wi93}. 
This approach made it possible to construct 
a large class of $(0,2)$ string vacua that is field theoretically 
more managable \cite{di94}. 
The starting point is a $(0,2)$ supersymmetric $U(1)$ gauge 
theory that leads in the Calabi-Yau phase to a $(0,2)$ $\sigma$ model 
described by the follwing geometric 
data: 
\begin{eqnarray}
0\longrightarrow V \longrightarrow \bigoplus^{r+1}_{i=1}{\cal O}(n_{i})
\stackrel{F}{\longrightarrow} {\cal O}(m)\longrightarrow 0	\label{VB}
\end{eqnarray}
is an exact sequence defining a stable bundle $V$ of rank $r$ 
on a complete intersection Calabi-Yau variety $M$ 
in a weighted projective space ${\Bbb P}_{w_{1},\ldots, w_{N+1}}$. 
In the following we will abbreviate the vector bundle data 
by $V_{n_1,\ldots,n_{r+1}}[m]$, where 
$n_{i}$ are positive integers and $F_{i}$ are homogeneous 
polynomials of degrees $m-n_{i}$ not vanishing simultaneously on $M$.
The choices $r=4$ and $r=5$ correspond to unbroken gauge 
groups $SO(10)$ and $SU(5)$, respectively.
As pointed out above these  geometric data are subject to 
constraints which in this case lead to the following equations:  
Let $M$ be the complete intersection of hypersurfaces of degrees $d_{j}$,
$j= 1,\ldots, N-3$. Then the Calabi-Yau condition 
for $M$, i.e. $c_{1}(M)=0$,
leads to:   
\begin{equation}
	\sum_{j=1}^{N-3}d_{j}-\sum_{i=1}^{N+1} w_{i}=0 .\label{1}
\end{equation}
Taking the defining sequence of $V$ into account, 
the condition $c_{1}(V)=0$  means that 
\begin{equation}
	           m-\sum_{i=1}^{r+1} n_{i}=0 .\label{2}
\end{equation}
The last condition, which comes from the cancellation of gauge anomalies, 
results in the quadratic diophantine equation
\begin{equation}
      m^{2}-\sum_{i=1}^{r+1} n_{i}^{2}=\sum_{j=1}^{N-3}d_{j}^{2}
                                       -\sum_{i=1}^{N+1}w_{i}^{2}.\label{3}
\end{equation}
For a Calabi-Yau hypersurface of degree $d$ the choice $m=d$ and 
$n_{i}=w_{i}$ solves the above 
equations. 
In the special case of $F_{i}=\partial_{i}W$, where $W$ is the
transversal polynomial that defines $M$ in ${\Bbb P}_{w_{1},\ldots, w_{5}}$, 
we have a $(2,2)$ model with gauge group $E_{6}$.
For a generic choice of $F_{i}$ the bundle $V$ will 
be a stable deformation of the extension of ${\cal T}_{M}$ by $\cal O$.
These models are therefore $(0,2)$ deformations of $(2,2)$ models (cf. 
\cite{di94} for more details).

\subsection{A series of solutions} 

Our aim is to find a generalization of the map from a (0,2) SCFT to a set of
vector bundle data that has been found in \cite{bl95}. In that paper the 
weights $n_i$ of a transverse CY 
hypersurface in weighted ${\Bbb P}^4$ correspond to the quintic 
(i.e. $n_i=1$ and $m=\sum n_i=5$) and it is these numbers that enter the 
definition of the
vector bundle, whereas the base manifold gets modified and turns out to be
the vanishing locus of two degree 4 polynomials in ${\Bbb P}_{1,1,1,1,2,2}^5$.
At first glance it may be surprising that it is
the base manifold and not the vector bundle data that changes. This is, 
however, 
in agreement with the fact that the discrete gauge symmetry that survives
the breaking of the $U(1)$ in the gauged linear $\s$ model is a $\IZ_m$
with $m$ defined in eq. (\ref{VB});
this symmetry should be identified with the $\IZ_m$ quantum symmetry 
\cite{va891} that comes with the GSO projection on the CFT side.

As we will see in our discussion of the simple current construction of the 
(0,2) model, the breaking of the left-moving SUSY is accomplished by a
twist that acts only on a Fermat factor of the tensor product constituting the
internal SCFT and on an SO(2) factor of the current algebra. 
This twist leads to a $\IZ_2$ projection in the NS sector, 
which explains the doubling of the last weight and the requirement
of an additional generator for the cohomology to discribe the contributions 
from the twisted sectors. This additional generator is obtained by increasing
the dimension of the ambient space and hence the codimension of the base 
manifold.
 
\del
We begin with the vector bundle data $V_{n_1,\ldots,n_5}[m]$ 
which are associated with the rational weights $q_i=n_i/m$ with 
$m=\sum n_i$ of a transversal CY hypersurface in weighted 
${\Bbb P}^4$ which is suggested by the ${\Bbb Z}_m$ quantum symmetry 
\cite{va891} of the corresponding $(0,2)$ LG model that should correspond
to the ${\Bbb Z}_m$ symmetry that comes with the generalized GSO 
projection in the Gepner model \cite{ge88}. From the example in \cite{bl95} 
we guess that the base CY
variety should be a complete intersection of codimension 2 in 
weighted ${\Bbb P}^5$. 
\enddel

We are thus lead to the ansatz $ V_{n_1,\ldots,n_5}[m]\to
{\Bbb P}_{n_1,\ldots,n_4,w_5,w_6}[d_1,d_2]$ for the base
manifold and vector bundle data, which we expect to work whenever 
there exists a transverse polynomial of degree $m$ in 
four fields $\Ph_i$ with weights $n_1,\ldots,n_4$ and with $K=m/n_5$ being
an odd integer (i.e. the internal CFT should be a tensor product with a
Fermat factor corresponding to a minimal model at odd level).
It is a non-trivial check on the existence of the desired series of maps
that there are {\it positive integer} solutions to the anomaly 
equations (\ref{1}--\ref{3}). Demanding that the quadratic equation (\ref{3})
factorizes we indeed find a unique (up to permutation of weights) general 
solution with $w_5=2n_5$, $2w_6=d_1=m-n_5$ and $d_2=(m+3n_5)/2$.
\del
We need to fulfill the constraints (\ref{1}),(\ref{2}) and(\ref{3})
for a codimension $2$ complete intersection $M$ and a rank 
$4$ stable bundle $V$. If the mechanism really works we 
expect a natural generalization of the 
\cite{bl96} solution with $n_i=1$, $d_i=4$ and $w_1=\ldots=w_4=1$, 
$w_5=w_6=2$.
This should work whenever, say, $m/n_5$ is an odd integer and there is
a transversal polynomial of degree $m$ that depends only on $\phi_1,\ldots,
\phi_4$. The existence of a solution of the above diophantine equations 
for arbitrary $n_1,\ldots,n_4$ is a non-trivial check for this picture.
Making the ansatz $w_i=n_i$ for $i=1,\ldots,4$ and playing around with the
equations we indeed find a general solution:$ V_{n_1,\ldots,n_5}[m]\to
{\Bbb P}_{n_1,\ldots,n_4,2n_5,\frac{m-n_5}{2}}[m-n_5,(m+3n_5)/2]$,
i.e. $w_i=n_i$ for $i<5$, $w_5=2n_5$, $w_6=(m-n_5)/2=d_1/2$ 
and $d_2=(m+3n_5)/2$.
In this way we obtain a map which associates to each consistent geometric data 
$V_{n_1,\ldots,n_5}[m]\to {\Bbb P}_{n_1,\ldots,n_4,n_5}[m]$
a new one $V_{n_1,\ldots,n_5}[m]\to {\Bbb P}_{n_1,\ldots,n_4,2n_5,
\frac{m-n_5}{2}}[m-n_5,(m+3n_5)/2]$.
\enddel
It is encouraging that the expected doubling of the Fermat weight comes about
automatically.
\del
, which is encouraging since an existence of any general integer
solution to
the quadratic anomaly cancellation condition is by no means guaranteed, 
and a doubling of the weight of the 5$^{th}$ homogeneous coordinate
appears to be just what we need for describing the ${\Bbb Z}_4$ 
quotient action in the Fermat factor (the twist has order 2 in the 
NS sector (see below)). 
The extra coordinate should somehow provide the
twisted sectors. We thus expect the identification to work at least if the
Fermat factor corresponds to a minimal model at odd level, i.e. 
$m/n_5\in2{\Bbb Z}-1$. 
This implies that $m-n_5$ is even so that all degrees are integer.
\enddel
Searching through the list of 7555 transverse weights \cite{nms,kl94} we find
3219 LG potentials with 4 fields and central charge $c=6(1+1/K)$
with $K\in2\IZ+1$ being an odd integer. 
220 of these have the property that all weights $n_i$ 
divide the degree $m$, so that they correspond to tensor products of minimal
models.


In the Calabi--Yau phase of our $\s$ model we may be confronted with 
singularities in both, the base manifold and the vector bundle \cite{di96}.
It is natural to try a toric approach to the resolution of these singularities.
A canonical Calabi-Yau variety can be realized
as a complete intersection in a Gorenstein  Fano toric variety in the
following way \cite{bo93,ba94,ba95}: Using the correspondence of reflexive 
polytopes and  
Gorenstein  Fano toric varieties we begin with a reflexive polytope 
$\Delta$ in a lattice of rank $n$ and construct its corresponding 
toric variety $X=X_{\Delta}$ \cite{ba94}. 
Next we consider a nef partition 
of the anticanonical divisor $-K_{X}=\sum^{N}_{\alpha=1}D_{\alpha}$ 
of $X$, which is a partition of $-K_{X}$ into a sum of nef Cartier divisors
$D_{j}=\sum_{\alpha\in I_{j}}D_{\alpha}$ with $\bigcup^{r}_{j=1}I_{j}$ 
being a partition of $\{1,\ldots,N\}$. 
If $\Delta_{j}$ are the polytopes associated to $D_{j}$ it follows that 
$\Delta=\Delta_{1}+\ldots +\Delta_{r}$ is their Minkowski sum. 
Now let $Y_{j}$ be a generic section of ${\cal O}_{X}(D_{j})$.
Then the complete intersection $\bigcap^{r}_{j=1}Y_{j}$ will be a canonical
Calabi-Yau variety of codimension $r$ in $X$.                  

To obtain a toric resolution of the singularities in the 
base manifold 
it is clear that we should choose $\D_i$ to be (subpolytopes 
of) the Newton polytopes corresponding to degree $d_i$ monomials in 
$\Ph_1,\ldots,\Ph_6$ with $i=1,2$. Obviously $\D_1+\D_2$ is then a subpolytope
of the Newton polytope corresponding to the weights $n_1,\ldots,n_4,w_5,w_6$,
but it is not clear under what conditions such a $\D$ will be reflexive and
it may be necessary to study the situation on a case by case basis.
There is no space to go into further details at this point \cite{mn96}.

\del
Having this in mind we can put our foregoing dicussion 
in the following geometrical setting. We begin with $V_{n_1,\ldots,n_5}[m]
\to {\Bbb P}_{n_1,\ldots,n_4,n_5}[m]$ that is associated to 
the $(2,2)$ model. Due to a theorem of Skarke
we can take as a reflexive polytope $\Delta_{4}$ the Newton polytope 
of degree $m$ monomials on ${\Bbb P}_{n_1,\ldots,n_4,n_5}$. 
After resolving the singularities $M$ will be a Calabi-Yau 
hypersurface in a toric variety which is a blowup of ${\Bbb P}_{n_1,
\ldots,n_4,n_5}$. 
If we do the same thing for $V_{n_1,\ldots,n_5}[m]\to 
{\Bbb P}_{n_1,\ldots,n_4,2n_5,\frac{m-n_5}{2}}[m-n_5,(m+3n_5)/2]$ which is
associated to the $(0,2)$ model we get a polytope $\Delta_{5}$. 
The above mentioned 
theorem, however, does not apply in this case and
we need a specific argument.  
\enddel


\section{Simple currents, modular invariants and orbifolds}

In this section we recall some elements of the classification of simple current
modular invariants, which will allow us to give an orbifold interpretation to 
the construction of \cite{bl95} and to derive explicit formulas for the 
spectra in terms of the {\it extended Poincar\'e polynomial} \cite{sc91,kr95}
(which is related to the elliptic genus).

A simple current $J$ 
is a primary field that has 
unique fusion products with all other primary fields, and thus 
decomposes the field content of a CFT into orbits of finite length
\cite{sc90r}. 
The order $N$ of 
$J$ is the length of the orbit of the identity, i.e. $J^N=\Id$.
Associativity and commutativity
of the fusion product imply that the simple currents form an abelian group,
which is called the center of the CFT. 
The important fact 
is that we can define a {\it monodromy charge}
\BE
	Q_J(\Ph)\equiv h_J+h_\Ph-h_{J\ex\Ph} ~~~\mod~ 1
\EE
that is conserved modulo 1 in operator products.
Since $e^{2\p i Q_J}$ is conserved 
in OPEs, a simple current is thus always
associated to a discrete $\IZ_N$ symmetry of the CFT 
and the center has a representation in terms of 
phase symmetries. 
The definition of $Q_J$ 
implies 
\BE
	Q_{J\ex K}(\Ph)	\equiv Q_J(\Ph)+Q_K(\Ph), 
\EE
so that $Q_{J^n}(\Ph)\equiv nQ_J(\Ph)$ and 
$Q_J$ is quantized in units of $1/N_J$. 

For some fixed subgroup $\cg$ of the  center 
that is generated by independent simple currents $J_i$ of order $N_i$
we use the notation $[\a]=\prod J_i^{\a_i}$ and $Q_i=Q_{J_i}$, where 
$\a_i$ are integers that are defined modulo $N_i$. 
Then we can parametrize the charges and conformal weights  of all
simple currents in $\cg$ in terms
of a matrix $R_{ij}$ \cite{ga91},
\BE
     R_{ij}= r_{ij}/N_i\equiv Q_i(J_j) \equiv Q_j(J_i),  
\EE     
\BE
     h_{[\a]}\equiv \2\sum_i r_{ii}\a^i-\2\sum_{ij}\a^iR_{ij}\a^j,
\EE
with $r_{ij}\in\IZ$.
If $N_i$ is odd we can choose $r_{ii}$ to be even. Then
all diagonal elements $R_{ii}$ are defined modulo 2 for both, 
even and odd $N_i$. Using the definitions of $Q$ and $R$ we obtain
\BE
     h_{[\a] \Ph}\equiv h_\Ph+h_{[\a]}-\a^iQ_i(\Ph), 
\EE
\BE     
     Q_i([\a] \Ph)\equiv Q_i(\Ph)+R_{ij}\a^j.
\EE
If $N_i$ is even then
$J_i$ can occur in a modular invariant only if $r_{ii}\in2\IZ$
(since $T$-invariance requires $h_a-h_{J_ia}\in\IZ$)%
\footnote{ ~
	This is related to the distinction between odd and even orders
	in the orbifold level matching conditions.
}.
In that case only a subgroup of the center that 
consists of simple currents $J_i$ whose parameters $r_{ii}$ all are even can 
contribute to a modular invariant.   
For such a group it	
can be shown 		
that the matrix
\BE \label X	
     M_{\Ph,[\a]\Ph}=\m(\Ph)\prod_i\d_\IZ\(Q_i(\Ph)+ X_{ij}\a^j\)
\EEL{M}
commutes with the generators $S$ and $T$ of modular transformations if $X$ is
properly quantized and $X+X^T\equiv R$ (for the diagonal $2X_{ii}\equiv R_{ii}$
must hold modulo 2) \cite{kr94}. In this formula $\d_\IZ(r)$ is 1 if $r\in\IZ$
and 0 otherwise, and $\m(\Ph)$ is the multiplicity of the primary field
$\Ph$ on its orbit.
Under certain 
assumptions 
(\ref X) can be shown to be 
the most general 
modular invariant that only relates primary fields on 
orbits of the center \cite{kr94,ga91}.

It is easy to see that the left and right chiral algebra extensions are given
by the kernels  $\ca_R\cong\mao{Ker}_\IZ X$ and $\ca_L\cong\mao{Ker}_\IZ X^T$,
respectively,
where $\mao{Ker}_\IZ M$ denotes the set of integer vectors $\a$ whose
product with the matrix $M$ is integer.
($\ca_R$ and $\ca_L$ have the same dimension, but need not be 
isomorphic \cite{kr94}.)

\subsection{Symmetry breaking and GSO } 	

						\def\su{{v}} \def\sp{{s}}

An important application of the chiral algebra extension mechanism 
that we just discussed is the Ramond/NS sector alignment in the 
superconformal tensor product, which can be understood as a simple current 
modular invariant because the supercurrent $J_v$ of any superconformal field 
theory is a simple current with $h=3/2$ and order 2.
Its monodromy charge is $0$ in the $NS$ sector and $1/2$ in the Ramond sector.
In the conventional tensor product
of two SCFTs we thus have the current $J_v^{(12)}=J_v^{(1)}\ex J_v^{(2)}$,
which has integer spin and can extend the chiral algebra. 
The monodromy parameter $R_{vv}$ is indeed 0 modulo 2, so that the 
modular invariant (\ref M) keeps only
states with integer charge w.r.t. $J_v^{(12)}$, i.e. both factors of a
field in  the tensor product must come from the same sector.

The GSO projection can be implemented in a similar way: In any $N=2$ SCFT the
Ramond ground state $J_s$ with maximal charge is a simple current 
(which implements the spectral flow; for details see \cite{kr95}). Therefore
$J_{GSO}=J_s\ex s$, with $s$ being the spinor of $SO(10)$, is a simple current
with spin 1 that extends the gauge group to $E_6$
and leads, after the string map, 
to space-time SUSY.

For a tensor product of $n$ SCFTs we have to put all bilinears in
the respective supercurrents into the chiral algebras \cite{sc91,alush}. 
In this case we have more freedom since the modular invariant is given by a 
$(n-1)\ex(n-1)$ matrix $X$ for which only the symmetric part has to
vanish. 
If we choose $E_{ij}=\2(X_{ij}-X_{ji})=0$ then $X\equiv0$ and we have
the maximal extension of the chiral algebras and complete alignment of $R$ and
$NS$ sectors for both chiral halves. If $E\neq0$ then some bilinears are
projected out and SUSY is broken on both sides. We can construct
models where the alignment is kept on the right-moving side if we have 
a larger center with additional simple currents of even order. This is one
possible mechanism to construct (0,2) models with gauge group $SO(10)$,
which in general is not extended to $E_6$ by the GSO projection
if the left-moving SUSY is broken.

The mechanism for the construction of (0,2) models that is most natural from 
the point of view of $\s$ models \cite{hu85,wi86}
is closely related to what we just discussed,
but also involves simple currents of the gauge group.
Since our aim is to construct a heterotic string via the string map applied
to the right-moving sector
we need to keep the $SO(10)\ex E_8$ on that side and thus have to 
use some asymmetric construction. 

The simple current way to break a symmetry is to start from building
blocks that belong to a smaller chiral algebra and to view the CFT
with broken symmetry as a modular invariant with `non-maximal' extension
of the chiral algebra. We therefore start with a 4-dimensional bosonic string 
with an internal $N=2$, $c=9$ SCFT and a current algebra 
$SO(2)\ex SO(8)\ex E_8$, which can be extended to $SO(10)\ex E_8$ by putting
the product $v\ex V$ 
of the vector currents of $SO(2)$ and 
$SO(8)$ into the chiral algebra, as we did for the Ramond sector alignment
above
(we denote the $SO(2)$ and $SO(8)$ representations 
by small and capital letters, respectively). 

If we avoid that extension on the right-moving side we can still 
construct a modular invariant heterotic string with the string map, but
the GSO projection on the left-moving side will in general extend 
$U(1)\ex SO(8)$ only to $SO(10)$, where the $U(1)$ factor is a combination of 
the $SO(2)$ and the $U(1)$ of the $N=2$ algebra. Non-abelian
gauge groups of smaller rank
are obtained by splitting $SO(10)$ into several $SO(2)$ factors.

\subsection{$SO(10)$ models}
Returning to the SO(10) model based on the quintic that was conjectured to
be equivalent to a Distler--Kachru model in \cite{bl95} we first
need to bring the modular invariant that enters the construction into the 
normal form (\ref X).
In addition to the alignment currents and the GSO projection there is only
one more simple current involved, namely the order 4 current 
$J_b={\bf1}\ex\Ph^{01}_K\ex s\ex {\bf1}$, i.e. the product of the identities 
in the CFT with $c=6(1+1/K)$ and in the $SO(10)$ factor with the primary field 
labeled by $\{l=0,q=K,s=1\}$ of the minimal model at level $k=K-2$ and the 
spinor of $SO(2)$. A simple calculation shows that
all monodromies among these currents vanish for odd $k$ except for 
$Q_b(J_A)=Q_A(J_b)=1/2$,
where $J_A={\bf1}\ex{\bf1}\ex v\ex V$ is the alignment current for $SO(2)$ 
and $SO(8)$.
$J_b$ thus indeed prohibits a simultaneous extension to $SO(10)$ on the left
and on the right. 
Note that $J_b^2={\bf1}\ex J_v^F\ex v\ex {\bf1}$ is the alignment current 
for the 
minimal model and the $SO(2)$, so that $J_b$ can be regarded as a square root 
of an alignment current. 

Since the conformal weight $h_b=(K-1)/2$ of $J_b$ 
depends on the level it is convenient to replace it by its product 
$J_B$ with the alignment current $J_b^2J_A$. 
This product has integer spin. 
In the resulting basis 
\del
Regarding these facts and inspired by the work of \cite{bl95,bl96} we take
the follwing diagonal modular invariant as our starting point. The building
blocks are a $N=2$ SCFT $\bf C$, a $N=2$ minimal model $\bf F$ at odd level 
$k(=K-2)$, a $SO(2)$ and a $SO(8)\times E_{8}$ current 
algebras\footnote{ For the sake
of concreteness we only consider the $SO(10)$ case.}. The linearly realized
part of the gauge group is therefore $SO(2)\times SO(8)$. 
For $K\in 4{\Bbb Z}+1$ the SCMI is obtained by the following simple currents:
\begin{eqnarray}
 J_{\scriptscriptstyle GSO} & = & J_{s}^{\scriptscriptstyle C}\otimes 
                   J_{s}^{\scriptscriptstyle F}\otimes 
                   J_{s}^{\scriptscriptstyle SO(2)}  
                   \otimes J_{s}^{\scriptscriptstyle SO(8)},\\ 
J_{\scriptscriptstyle A}    & = & {\bf 1}\otimes {\bf 1}\otimes 
                   J_{v}^{\scriptscriptstyle SO(2)}  
                   \otimes J_{v}^{\scriptscriptstyle SO(8)},\\                  J_{\scriptscriptstyle B}  & = & {\bf 1}\otimes (J_{s}^
                   {\scriptscriptstyle F})^{\scriptscriptstyle K}\otimes
                   J_{s}^{\scriptscriptstyle SO(2)}\otimes {\bf 1},\\
J_{\scriptscriptstyle C}   & = & J_{v}^{\scriptscriptstyle C}
                     \otimes{\bf 1}\otimes {\bf 1}  
                     \otimes J_{v}^{\scriptscriptstyle SO(8)}.
\end{eqnarray} 
\enddel
\BEA
	J_{\scriptscriptstyle GSO} &=& J_{s}^{\scriptscriptstyle C}\ex 
		J_{s}^{\scriptscriptstyle F}\ex s\ex S\\
	J_{\scriptscriptstyle A}    & = & {\bf 1}\ex{\bf 1}\ex v\ex V\\
	J_{\scriptscriptstyle B}    & = & {\bf 1}\ex 
		(J_{s}^{\scriptscriptstyle F})^K\ex s\ex (V)^{K-1\02}\\
	J_{\scriptscriptstyle C}    & = & J_{v}^{\scriptscriptstyle C}\ex 
		{\bf 1}\ex{\bf 1}\ex V
\EEA
we find a monodromy matrix $R$ whose only non-vanishing entries are 
$R_{AB}=R_{BA}=1/2$.

\del
\begin{center}
\begin{tabular}{||c||c|c|c|c\TVR52||} \hline\hline
$R$	& $J_{\scriptscriptstyle GSO}$& $J_{\scriptscriptstyle A}$ & 
$J_{\scriptscriptstyle B}$	& $J_{\scriptscriptstyle C}$ \\ \hline\hline
$J_{\scriptscriptstyle GSO}$	& 0	& 0	& 0 & 0 \\ \hline
$J_{\scriptscriptstyle A}$	& 0	& 0	& $\2$	& 0 \\ \hline
$J_{\scriptscriptstyle B}$	& 0 & $\2$ & 0 & 0 \\ \hline
$J_{\scriptscriptstyle C}$	& 0	& 0	& 0	& 0 \\ \hline\hline
\end{tabular}
\end{center}
Choosing the appropriate torsions we get the following $X$ matrix
\begin{center}
\begin{tabular}{||c||c|c|c|c\TVR52||} \hline\hline
$X$	& $J_{\scriptscriptstyle GSO}$& $J_{\scriptscriptstyle A}$ & 
$J_{\scriptscriptstyle B}$	& $J_{\scriptscriptstyle C}$ \\ \hline\hline
$J_{\scriptscriptstyle GSO}$	& 0	& 0	& 0 & 0 \\ \hline
$J_{\scriptscriptstyle A}$	& 0	& 0	& $\2$	& 0 \\ \hline
$J_{\scriptscriptstyle B}$	& 0	& 0	& 0 & 0 \\ \hline
$J_{\scriptscriptstyle C}$	& 0	& 0	& 0	& 0 \\ \hline\hline
\end{tabular}
\end{center}
For $K\in 4{\Bbb Z}-1$ we get the same $R$ and $X$ if we modify 
the simple current $J_{\scriptscriptstyle A}$
such that $\bf 1$ in the last factor is replaced by 
$J_{v}^{\scriptscriptstyle SO(8)}$. 

One can easily see that 
$J_{\scriptscriptstyle A} \notin {\cal A}_{L}$ but 
$J_{\scriptscriptstyle A} \in {\cal A}_{R}$
and $J_{\scriptscriptstyle B} \notin {\cal A}_{R}$ but 
$J_{\scriptscriptstyle B} \in {\cal A}_{L}$ . It is this ${\Bbb Z}_{4}$ twist
which is responsible for the breaking of the left moving supersymmetries.
The gauge group is extended to $SO(10)$ by GSO projection. The increase in the 
rank comes from a combination of the $U(1)$s in $\bf C$ and $\bf F$ and 
an extra $SO(2)$'s at level one that fill up the central charge.     
\enddel
 
We now need to choose discrete torsions in such a way
that $J_{GSO}$ and the alignment
currents $J_A$, $J_B^2$ and $J_C$ all are in the right-chiral algebra, i.e.
all respective columns of $X$ should vanish. This uniquely fixes the 
anti-symmetric part of $X$ with 
$X_{AB}=1/2$ being the 
only non-vanishing
entry. Without further calculation we can thus conclude that this must be 
the invariant that enters the model of \cite{bl95} in the special case of the
quintic.
It is straightforward to derive a formula for the numbers of $SO(10)$
representations for general (0,2) compactifications of this type in terms of
the extended Poincar\'e polynomial of the internal CFT \cite{mn96}.
What remains to be done is to check the matching of these date with the
conjectured $\s$-model twin and to provide further evidence that the (0,2)
CFT indeed is in the same moduli space.

\section{Conclusion}

We used an ansatz for the $(0,2)$ $\sigma$ model data that is inspired by the 
orbifold interpretation of a `prototype model' \cite{bl95} 
to find a large class of solutions to the anomaly cancellation conditions.
The resulting analogies between the coordinate ring structure on the 
geometrical side and the orbifold selection rules on the CFT side indeed 
support the conjecture that this provides a series of identifications
among apparently different constructions of (0,2) compactifications.

Only a minimal model factor of the internal $c=9$ SCFT
takes part in the symmetry breaking mechanism, so that we can use 
arbitrary $N=2$ SCFTs with $c=6(1+1/K)$ to construct `hybrid models',
whose numbers of non-singlet massless representations can be computed in terms
of their extended Poincar\'{e} polynomials \cite{kr95}.

On the sigma model side 
the problem of resolving singularities provides a number of
new mathematical challenges \cite{di96,mn96}; the Landau--Ginzburg side of the 
triality has been studied for a closely related class of models in the recent 
paper \cite{bl961}. 
Models with other gauge groups like $SU(5)$ can be
constructed following the same lines.

\del
Starting from the anomaly cancellation condition we have found 
a large class of consistent $(0,2)$ $\sigma$ model data. 
Using the simple current methods we have constructed exact CFTs which we 
`guess' to match with these data. 
As we have seen the $\bf C$ part of the 
internal sector didn't take part in the symmetry breaking therefore one
can take a `general' $N=2$ SCFT as the $\bf C$ part and use the extended 
Poincar\'{e} polynomial to calculate the number of nonsinglet massless 
states. It is desirable to have explicit formulas
for the number of $\bf 16$s, $\bf \overline{16}$s and $\bf 10$s in terms of
extended Poincar\'{e} polynomial \cite{kr95}. On the other hand, it would be of
interest to study the sigma model side 
of `$(0,2)$ triality' for the models considered here in more details. 
The problem of resolving the singularities in this 
context leads to some remarkable phenomena and 
opens the way to new mathematical possibilities \cite{di96}. 
Last but not least it would be interesting to study the phenomenological
consequences of the models with other gauge groups, say $SU(5)$, which can be
constructed following the same lines as discussed above. 
\enddel

{\it Acknowledgements.} This work has been supported by the 
	{\it Austrian Research Funds} FWF under grant Nr. { P10641-PHY}.

\bye